Effects of Yttrium doping on Oxygen Conductivity in Ba (Fe, Co, Zr, Y) $O_{3-\delta}$ Cathode Materials for Proton Ceramic Fuel Cells

Chiyoung Kim[a], Ryan Jacobs[a], Jack H. Duffy[b], Kyle S. Brinkman[b], Harry W. Abernathy[c], Dane Morgan[a]

a. Department of Materials Science and Engineering, University of Wisconsin – Madison, Madison, Wisconsin 54706, United States
b. Department of Materials Science and Engineering, Clemson University, Clemson, South Carolina 29634, United States
c. National Energy Technology Laboratory, United States Department of Energy, Morgantown, West Virginia 26507, United States

**Abstract**

Proton ceramic fuel cells (PCFCs) offer enhanced efficiency at lower operating temperatures compared to solid oxide fuel cells. However, their performance is often limited by cathodic oxygen reduction reaction (ORR) kinetics. The $BaCo_xFe_yZr_zY_{1-x-y-z}O_{3-\delta}$ (BCFZY) series of perovskite materials have been extensively investigated and shown to be highly promising as triple-conducting air electrode materials for PCFCs. However, fundamental knowledge of the role of acceptor dopants, such as Y, on oxygen conductivity remains unclear. In this study, we investigate the effects of Y doping on the oxygen conductivity of BCFZY materials with varying Y content: $BaFe_{0.4}Co_{0.4}Zr_{0.2}O_{3-\delta}$ (BCFZ), $BaFe_{0.4}Co_{0.4}Zr_{0.1}Y_{0.1}O_{3-\delta}$ (BCFZY0.1) and $BaFe_{0.4}Co_{0.4}Y_{0.2}O_{3-\delta}$ (BCFY). Oxygen conductivity was analysed in terms of the product of oxygen tracer diffusivity and oxygen concentration. Ab initio molecular dynamics (AIMD) simulations were used to calculate oxygen tracer diffusivity and its migration energy, while oxygen defect concentrations were estimated from reference data. Our results indicate that Y doping slightly decreases oxygen conductivity from BCFZ to BCFZY0.1 with values of 337 ± 105 mS/cm and 203 ± 81 mS/cm at 500 °C, respectively, with corresponding activation energies of 0.155 ± 0.06 eV and 0.172 ± 0.006 eV. In contrast, BCFY exhibits a significantly lower conductivity (99 ± 40 mS/cm) and a higher activation energy of 0.261 ± 0.006 eV than the more Y-poor materials. A comparison with experimental data shows that the computed conductivities are much higher and have more Arrhenius-like behavior in their temperature dependence than available experimental results, suggesting that microstructural effects such as grain boundaries likely play a key role in governing oxygen transport in experiments. Consistent with this hypothesis, a series circuit model of our calculated bulk conductivity and fitted grain boundary transport parameters resulted in semi-quantitative agreement with experimental conductivities and grain boundary behaviour consistent with that typically observed in oxides. These findings provide insights into the defect chemistry and ORR kinetics of BCFZY materials, offering guidance for optimizing cathode performance in future PCFC applications.

# 1. Introduction

Solid oxide fuel cells (SOFC) are promising renewable energy devices, known for their high efficiency and ability to generate energy using a variety of fuel sources (e.g., $H_2$, $CH_4$, and other hydrocarbons).[1,2] However, SOFCs typically require high operating temperatures of roughly 800 °C for adequate power generation. Such high operating temperature results in severe challenges, including the necessity for high-cost interconnect materials, accelerated materials degradation and higher materials replacement costs.[3] Proton ceramic fuel cells (PCFCs) have the potential to overcome these issues because they can operate at considerably lower temperatures (400-650 °C) than SOFCs.[4,5] This lower temperature operation is primarily enabled by the faster transport of protons compared to oxygen ions in the fuel cell electrolyte.[6] Power density (units of $W/cm^2$), which is one marker of performance of PCFCs, is considered the most critical parameter for determining stack cost. Higher-performing cells can deliver the same power output with fewer units, thereby reducing materials usage and stack assembly costs. [3] Similar to SOFCs, limited oxygen reaction reduction (ORR) kinetics of the cathode at reduced temperature often limits the performance of PCFCs. Gaining an improved understanding of materials properties dictating cathode ORR, such as oxygen conductivity, is one of the prerequisites for designing improved high power density PCFC systems.[7,8]

The $BaCo_xFe_yZr_zY_{1-x-y-z}O_{3-\delta}$ series of perovskite materials (henceforth BCFZY perovskites) were designed as triple conducting materials for PCFCs, capable of ample conduction of oxygen, hydrogen (in the form of protons) and electrons at elevated temperature.[9,10] BCFZY can now be considered a model benchmark material from which performance comparisons of new promising triple conducting materials are made. [4,8,11] The design of BCFZY triple conductors can be traced back to successive doping of the $BaZrO_3$ parent material. Y-doped $BaZrO_3$ perovskite is a well-known mixed ionic conductor exhibiting high proton and oxygen ion conductivity and very low electronic conductivity, making it an effective electrolyte material.[12] Separately, doping of redox active transition metals such as Co and Fe in $BaZrO_3$ to create, e.g., $BaCo_{0.4}Fe_{0.4}Zr_{0.2}O_{3-\delta}$ (BCFZ) significantly increased its electronic conductivity and ORR catalytic activity, producing a high-performing SOFC cathode material.[13] To improve the performance of BCFZ for PCFCs, research on proton uptake kinetics, surface oxygen reactions and structural stability was conducted.[14–19] Numerous studies introduced Y dopants in place of Zr in BCFZ materials, creating commonly-explored compositions like $BaCo_{0.4}Fe_{0.4}Zr_{0.1}Y_{0.1}O_{3-\delta}$ (BCFZY0.1) and the Y-rich $BaCo_{0.4}Fe_{0.4}Y_{0.2}O_{3-\delta}$ (BCFY).[6,20–23] It was expected that Y doping could help increase proton uptake by producing additional oxygen vacancies and localizing electrons on oxygen, similar to the effects in Y-doped $BaZrO_3$[13], and optimal Y doping in BCFZY materials produces a significant improvement on the performance of BCFZ. [13] More specifically, when comparing BCFZ to BCFZY0.1, the impact of Y doping results in a reduction of the area-specific resistance (ASR) by up to an order of magnitude at 500 °C, a sixfold increase in proton conductivity at 550 °C, while its oxygen conductivity at 550 °C is threefold lower. [6,24,25] Taken together, this needs to be understood better, as Y content may introduce trade-offs in the kinetics of different carriers (protons vs. oxide ions) under the relevant $pO_2/H_2O$ and temperature.

BCFZY materials are today considered as one of the most promising cathode materials in PCFC system due to their generally exceptional ability to facilitate ORR and concurrently transport electrons, oxygen ions and protons. Despite the significant effect of Y doping in BCFZY systems and numerous studies demonstrating the efficacy of BCFZY as a PCFC cathode material, understanding of how Y doping affects the oxygen conductivity of BCFZY systems remains limited. Previous research has experimentally measured oxygen conductivity but lacks a clear, in-depth understanding of the underlying mechanisms, particularly regarding the intrinsic bulk conductivity. Moreover, the role of microstructure in determining the overall performance of BCFZY systems remains largely unexplored, despite its importance for the practical design of fuel cells.

This lack of understanding, in turn, limits our ability to rationally design and optimize other triple conducting materials. In this study, we investigated the magnitude and mechanisms of how Y doping affects the oxygen conductivity of BCFZY materials. The key areas we focus on are: (i) how Y doping affects the oxygen diffusion (and therefore conductivity) through changes in vacancy content and migration energy, with a particular effort to understand why activation energy of oxygen conductivity (technically $\sigma T$) seems to reach its minimum near Y = 0.1, and (ii) what is the origin of the approximately 0.7 eV activation energy for conductivity (technically $\sigma T$) and why does it deviate from Arrhenius behavior in some experiments. [25,26] For oxygen diffusion, we used density functional theory (DFT) and *ab initio* molecular dynamics (AIMD) calculations to observe the effect of Y doping on the oxygen self-diffusion coefficient. For oxygen concentration, we used experimental data from previous study to fit our target condition. [17] Finally, we calculated oxygen conductivity and compared with experimental data from the previous work of Duffy et al.[25] Overall, we find that Y doping in BCFZ leads to a 25 percent decrease in oxygen conductivity for BCFZY0.1 and a 69 percent decrease for BCFY at 500 ℃, relative to BCFZ. The reduction in oxygen conductivity is primarily attributed to a decrease in oxygen diffusivity due to increased migration energies. This result is consistent with experimental results, suggesting that improvements in ORR performance with Y doping are not from faster oxygen transport, and may instead be due to increases in proton uptake or proton diffusivity, the analysis of which will be addressed in a separate study. Finally, we obtained migration energy ($E_{mig}$) for the oxygen diffusion in different BCFZY materials. We compared our conductivity results, which model a single crystal, to available polycrystalline experiments. We discuss discrepancies in terms of possible microstructural effects and propose that the discrepancies may be due to slow oxygen diffusion in grain boundaries.

## 2. Methods
### 2.1. *Ab initio* molecular dynamics (AIMD) method for calculating oxygen diffusivity

To explore the dynamics of oxygen diffusion in the BCFZY materials, AIMD calculations were carried out with DFT using the Vienna ab initio simulation package (VASP).[27] To prepare each BCFZY composition for AIMD simulations, the structures were first fully relaxed (with symmetry constraints removed) to obtain their most stable ground-state configurations. Using the resulting relaxed cell volume, the cell dimensions were then adjusted to create a pseudocubic geometry, in which the lattice parameters were set a = b = c, while preserving the total volume. A subsequent relaxation of only the ionic positions was then performed within this fixed volume pseudocubic cell. For this step, a 4×4×4 Monkhorst-Pack k-point grid was applied to the Brillouin zone of a 2×2×2 supercell (40 atoms) derived from the cubic perovskite primitive unit cell (5 atoms).[28] AIMD calculations were subsequently carried out using gamma-point-only sampling of reciprocal space. The calculations used the Perdew, Burke and Ernzerhof (PBE) generalized gradient approximation (GGA) for the exchange-correlation functional, with a Hubbard U correction applied within the PBE-type pseudopotentials.[29,30] The U values used in the PBE + U method were consistent with those adopted by the Materials Project: U = 5.3 eV for Fe and 3.32 eV for Co.[31] Core electrons were treated using projector augmented-wave (PAW) pseudopotentials, and the calculations were performed with a plane-wave energy cutoff of 500 eV.[32] An NVT ensemble was employed to maintain a constant number of atoms, volume, and temperature throughout the AIMD simulations, which were conducted with a time step of 1 femtosecond (fs) for a total time of 100 ps. The system temperature was set to 1000 K, 1333 K, 1666 K, and 2000 K using a Nose-Hoover thermostat.[33,34] The oxygen non-stoichiometry in the BCFZY systems was established based on the oxidation states of the active redox transition metals, namely Fe and Co, which were all assumed to be in the 3+ oxidation state, while Ba, Zr, Y, and O were all assumed to be redox inactive and maintain constant oxidation states of 2+, 4+, 3+ and 2-, respectively. The initial supercells consisted of a 2×2×2 expansion of the 5-atom $BaZrO_3$ perovskite unit cell, for a total of 40 atoms. The stoichiometric supercell compositions of BFCZ, BFCZY0.1, and BFCY consist of 8 Ba, 3 Fe, 3 Co, and 24 O (note, the Fe and Co concentrations are 37.5%, rather than 40% of the B-site, due to supercell size limitations). For BCFZ, in the supercell there are 2 Zr atoms, while for BCFZY0.1 there are 1 Zr and 1 Y and for BCFY there are 2 Y atoms. To maintain Fe and Co in their fixed 3+ oxidation states, oxygen vacancies were introduced to approximate overall charge neutrality in the supercells. Electrons were added or removed as necessary to ensure that Fe and Co, as well as Ba, Zr, Y and O, retained their assigned oxidation states of 3+. 2+, 4+, 3+ and 2-, respectively. Specifically, three oxygen vacancies were introduced for BCFZ and four for BCFZY0.1 and BCFY, corresponding to oxygen non-stoichiometries ($\delta$) of 0.375 for BCFZ and 0.5 for BCFZY0.1 and BCFY, respectively.

To predict diffusion, we used AIMD on the above described supercells and the relationship between oxygen ion diffusivity and vacancy concentration, $D_O = \Lambda D_{vac}$, where $\Lambda = c_{vac}/c_O$ ($c_{vac}$ and $c_O$ are the concentrations of vacancies and oxygen ions). We first calculated $D_O$ from AIMD simulations in the supercells with the above vacancy concentrations and then derived $D_{vac}$ using the Λ values corresponding to the supercell composition used in AIMD (Λ = 0.14, 0.2, and 0.2 for BCFZ, BCFZY0.1, and BCFY, respectively). We then calculated δ values from the reference data (described more below) and calculated predicted Λ values for the true materials. These Λ values were multiplied by $D_{vac}$ to yield the $D_O$ predicted for the materials at their true experimental vacancy concentrations. A complete list of the values of the relevant variables associated with the diffusion calculations is provided in Table S1.

The AIMD calculations were performed with six different B-site configurations. The six different B-sites were determined based on the distance between the acceptor dopants, Zr and Y. First, Zr and Y were arranged to obtain three different distances for each of the Zr-Zr, Zr-Y, and Y-Y pairs. Then, two random configurations of Co and Fe atoms were made for each of the Zr and Y arrangements, resulting in six different B-site cation configurations. For the magnetic state determination, a ferromagnetic ordering was assumed and used as a practical approximation to the paramagnetic state which occurs in the real system at fuel cell operating temperatures. [35] The ferromagnetic ordering was set by configuring all the Fe and Co atoms in the high-spin state with parallel coupling and keeping the total magnetic moment of the supercell fixed. This total moment sets the electron configurations as $Co^{3+}$ ($d^6$) and $Fe^{3+}$ ($d^5$) and fixed total magnetic moment as 27 $\mu_B$ per the ideal 2×2×2 supercell.

## 2.2. Diffusivity calculation

We employed the Time-Averaged Mean Square Displacement (TAMSD) method to compute the self-diffusivity of oxygen in the BCFZ, BCFZY0.1 and BCFY systems.[35,36] Figure S1 shows the tracer diffusivity of oxygen atoms and the migration energy for oxygen diffusion in BCFZY systems with six different B-site configurations, where the oxygen diffusion was calculated by averaging these results to obtain a representative solid solution diffusivity. (Figure 1) Note that this averaging is consistent with a model where these pathways are all available to the diffusing oxygen in parallel (if they are available sequentially then one would average inverses of each to get the average inverse). Configurations 1 and 2 correspond to the shortest Zr-Zr, Zr-Y, and Y-Y distances, configurations 3 and 4 represent medium distances, and configurations 5 and 6 correspond to the longest distances within the 2×2×2 supercell of the BCFZY systems. The Standard Error of the Mean (SEM) of the oxygen diffusivity in Figure 1 was calculated from the variation among six different B-site configurations. The uncertainty in the oxygen migration energy ($E_{mig}$) was obtained by error propagation of the SEMs through the linear fit of the natural logarithm of diffusivity versus inverse temperature, using weighted least-squares regression, as implemented in the Levenberg-Marquardt algorithm (curve_fit in scipy). [37]

## 2.3. Estimating oxygen concentration from experimental non-stoichiometry data

To determine the oxygen concentration (3 – δ) in BCFZ, BCFZY0.1, and BCFY, we used oxygen non-stoichiometry (δ) data reported by Zohourian et al. [17] across different temperatures and oxygen partial pressures (100 ppm, 1%, and 100% $O_2$). For each temperature, δ was linearly fitted as a function of $\ln(pO_2)$, using ordinary least-squares regression (np.polyfit in NumPy). [38] The fitted relation was used to extrapolate δ at 0.21 atm $O_2$, corresponding to ambient air conditions. (Figure S2) This extrapolated value was then used to calculate the oxygen concentration under our target conditions. The Python script used for the fitting and extrapolation can be found in the Data Availability Statement section.

## 2.4. Universal Machine Learning Interatomic Potentials

To explore the influence of system size on oxygen diffusion, molecular dynamics (MD) simulations were conducted utilizing the Universal Machine Learning Interatomic Potential (U-MLIP) 3-body Materials Graph Network (M3GNet) [39] to decrease computational demands vs. full AIMD. These MD simulations were carried out using the atomic simulation environment (ASE) python library. The study included MD calculations for the same 2×2×2 supercells (37 atoms) used in our AIMD studies, as well as for larger 4×4×4 (296 atoms), 8×8×8 (2368 atoms), 12×12×12 (7992 atoms) and 16×16×16 (18944 atoms) supercells. Each simulation ran up to 100 picosecond (ps) with a time step of 1 fs under an NVT ensemble. The observed oxygen diffusion mechanisms in all simulation videos were identical, showing substitutional oxygen diffusion through oxygen vacancies. The calculated diffusion coefficients show some variation between smaller cells ($E_{mig}$ = 0.19 ~ 0.32 eV for 2×2×2 to 8×8×8), but stabilize at 0.30 ~ 0.31 eV for 12×12×12 and 16×16×16 systems, indicating that the results are effectively converged by the 12×12×12 supercell (Figure S3). The converged supercells exhibit a small drop of diffusion coefficient of about 0.1 ~ 0.2 log units compared to the 2×2×2 supercell. These findings suggest that cell size effects are present but relatively minor for both $D_O$ and $E_m$, and may explain part of the discrepancy between the model and the experimental oxygen conductivity data (Figure 3), though they are unlikely to be the primary cause.

## 3. Results and Discussion
### 3.1. Effect of Y doping
#### 3.1.1. Effect of Y doping on oxygen concentration in BCFZY materials

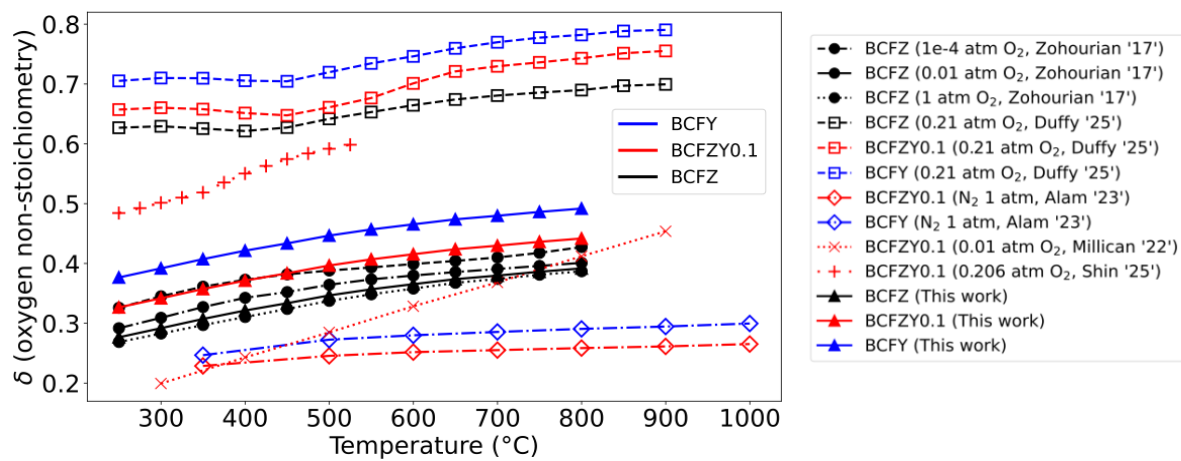

Figure 1. Temperature dependence of oxygen non-stoichiometry, δ, for BCFZ, BCFZY0.1, and BCFY. Unified color scheme across all traces: BCFZ = black, BCFZY0.1 = red, BCFY = blue. Literature datasets: Zohourian et al. [17] - BCFZ at $pO_2$ = $10^{-4}$, $10^{-2}$, 1 atm (black dashed, filled circles); Duffy et al. [40] - BCFZ/BCFZY0.1/BCFY at $pO_2$ = 0.21 atm (dashed, open squares, colored by composition); Alam et al. [42] - BCFZY0.1 and BCFY in $N_2$ at 1 atm (solid, open diamonds); Millican et al. [43] - BCFZY0.1 at $pO_2$ = $10^{-2}$ atm (red dotted, "×"). Shin et al. [41] - BCFZY0.1 at $pO_2$ = 0.206 atm (red dotted, "+"). This work: solid lines with filled triangles, using the same color scheme.

Oxygen concentration $(3 - \delta)$ in the BCFZY materials were fitted using data from previous work, as discussed in Sec. 2.3. [40] The $\delta$ values vs. temperature for the three Y concentrations are shown in Figure 1. Oxygen concentration decreases with increasing temperature at constant $pO_2$ (0.21 atm), which is expected considering that entropy stabilizes the gas with increasing temperature, removing oxygen from the system. Y doping increases oxygen vacancies ($\delta$), which is also expected because its lower 3+ valence comparing to 4+ Zr atoms effectively oxidizing the system which leads to compensating reductive anion vacancies.

The oxygen vacancy concentration ($\delta$) used in our oxygen conductivity calculations (0.37 in BCFZ at 600 °C) is significantly lower than the largest value reported in the literature (0.66 in BCFZ at 600 °C, Duffy et al. [40]). This suggests that the assumption of just reducing Fe may not be totally correct at these high temperatures and the real system at least partially reduces Co. [41] Furthermore, it indicates that the experimental system contains a significant amount of reduced 3+ or 2+ Fe (and maybe Co) at higher temperatures. However, other literature reports provide comparable values for the oxygen vacancy concentration (δ): thermogravimetric analysis (TGA) indicates δ = 0.33 for BCFZ at 250 °C [17]; δ = 0.48 for BCFZY0.1 at 250 ° [41]; and, using the same method, δ = 0.23 for BCFZY0.1 and δ = 0.33 for BCFY at 350 °C [42]. In addition, a study employing DFT-based oxygen vacancy prediction model estimated δ = 0.20 for BCFZY0.1 at 250 °C. [43] These literature values are in closer agreement with the value we adopted from Zohourian et al. [17] than with those reported by Duffy at al. [40] Considering the spread in results shown in Figure 1 it seems that a consensus on precise values for δ has yet to emerge.

The differences in reported δ across studies can arise from both synthesis conditions and how the absolute δ reference was defined, and consequently how Δδ was computed. Zohourian et al. [17] and Alam et al. [42] sintered at 1200 °C, Shin et al. sintered at 1250 °C. whereas Duffy et al. [40] used 1275 °C. Such changes might affect the variation in microstructure and stoichiometry of cations. For δ determination, Zohourian et al. [17] established reference δ (not explicitly reported) by fully reducing BCFZ in 7% $H_2/N_2$ at 1100 °C (end phases verified by X-ray diffraction (XRD)) and then converted thermogravimetric mass changes at controlled $pO_2$ to calculate δ(T, $pO_2$). Duffy et al. [40] referenced room-temperature neutron powder diffraction (NPD)/XRD Rietveld oxygen occupancies to set δ = 0.624 for BCFZ, δ = 0.654 for BCFZY0.1, and δ = 0.702 for BCFY, then used TGA in air to obtain Δδ(T). Shin et al. [41] anchored δ = 0.531 (for BCFZY0.1) to neutron diffraction (ND)-refined occupancies for a sample equilibrated at 250 °C in dry air (then quenched) and measured Δδ by TGA. Alam et al. [42] fixed reference δ = 0.225 (for BCFZY0.1) and δ = 0.253 (for BCFY) by Mohr's-salt/$K_2Cr_2O_7$ redox titration (from the average Co/Fe valence) and tracked Δδ by TGA under $N_2$. It is likely that differences in the chosen reference δ have a larger impact on discrepancies between studies than Δδ, since the change in δ between low and high temperatures (~250–900 °C) is similar among the studies (~0.1 per $ABO_3$ formula unit; Figure 1). Given these uncertainties we use data from Zohourian et al. [17] as a reasonable estimate.

### 3.1.2. Effect of Y doping on oxygen diffusivities in BCFZY materials

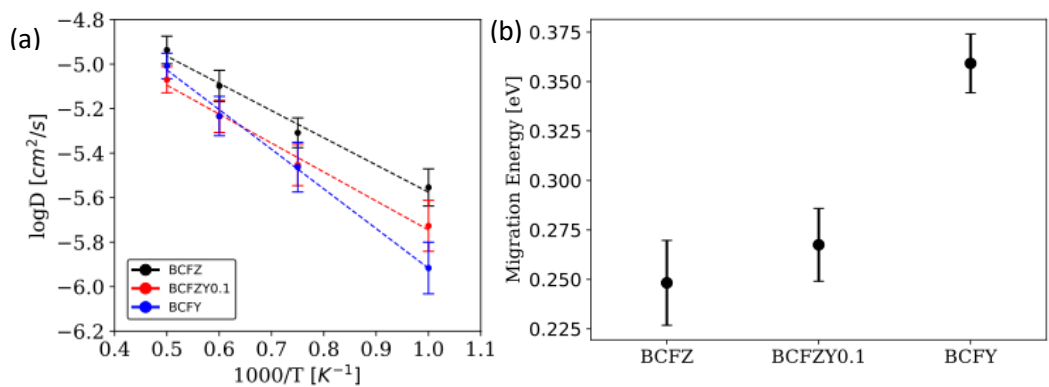

Figure 2. (a) Tracer diffusivity and (b) $E_{mig}$ of oxygen atom in BCFZ, BCFZY0.1 and BCFY. D values are determined from AIMD and scaled to values expected for thermodynamically equilibrium vacancy concentrations (see Sec. 2.1).

Oxygen atoms diffuse through a vacancy-mediated mechanism, where they hop between adjacent oxygen vacancies in the perovskite lattice [44], and oxygen diffusion can be studied by direct AIMD simulation of supercells with oxygen vacancies. Using AIMD, we calculated the oxygen tracer diffusion coefficients ($D_O$) and associated migration energies for BCFZ, BCFZY0.1 and BCFY (see Sec. 2.1 and 2.2), and the results are shown in Figure . From Figure (a) it was observed that introducing successive amounts of Y into BCFZ to form BCFZY0.1 and BCFY leads to a reduction in $D_O$ as Y doping is increased. Note that this reduction occurs despite Y doping increasing the oxygen vacancy concentration, which would normally be expected to enhance oxygen-conduction pathways, suggesting that the effect is likely due to changes in the migration energies of the oxygen. Several studies hypothesize that dopant-generated oxygen vacancies can associate with dopant cations, forming vacancy–dopant complexes that trap vacancies and thereby reduce the number of mobile defects and the ionic conductivity. [45–50] Such an association is very natural to hypothesize given the opposite Kröger–Vink charges on the vacancies and dopants and the observed decrease in $D_O$ and the higher migration energy with increasing Y are compatible with defect–dopant coupling. However, the AIMD in this work and our prior DFT study (Duffy et al. [40]) on BCFZY materials did not observe an O-vacancy–Y attraction but instead some level of repulsion. One possibility is that a vacancy repulsion from Y is driven by strong Y–O bonding (this was suggested by our earlier DFT, Duffy et al. [40]). This repulsion could impede vacancy motion and lead to an overall increase in the effective migration energy for oxygen, consistent with our AIMD observations. However, further study is required to establish more robustly the dominant mechanisms for how Y increases migration energetics. From the AIMD simulations, the migration energy of oxygen diffusion in BCFZ is $0.25 \pm 0.02$ eV, in BCFZY0.1 is $0.27 \pm 0.02$ eV and in BCFY is $0.36 \pm 0.01$ eV. Thus, in BCFY the migration energy for oxygen diffusion is significantly higher than in both BCFZ and BCFZY0.1. Overall, increasing Y content decreases the oxygen diffusivity and raises the migration energy for oxygen in the case of Y = 0.2 (BCFY) and appears to have similar but much a smaller effect for Y = 0.1, although the effect on $D_O$ and migration energy are within our uncertainties.

### 3.1.3. Effect of Y doping on oxygen conductivities in BCFZY materials

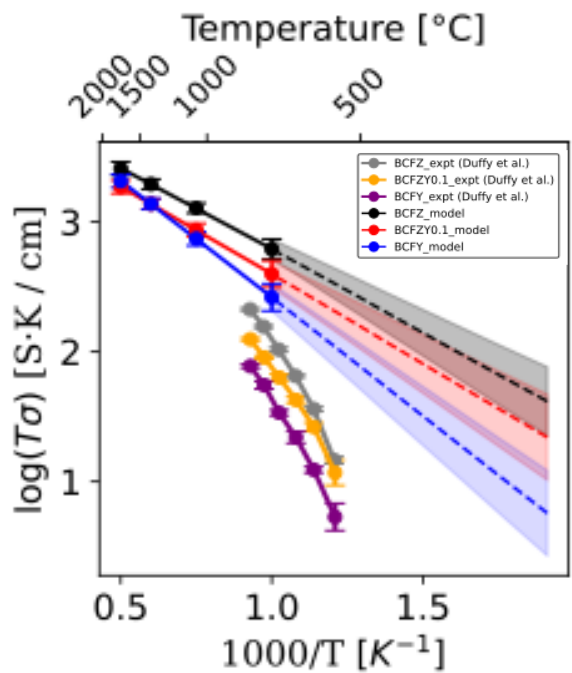

Figure 3. Oxygen conductivity ($\sigma_O$) of BCFZ, BCFZY0.1 and BCFY with experimental references.[25,26] Below 1000 K, conductivity was extrapolated, with error bars reflecting uncertainty propagated solely from the diffusivity ($D_O$), excluding any uncertainty from DFT.

To calculate oxygen conductivities ($\sigma_O$) in the BCFZY materials, the Nernst-Einstein relation [51] was used:

$$\sigma_O = \frac{Z^2 e^2 c_O D_O}{kT} \quad \text{(Equation 1)}$$

where the oxygen concentration ($c_O$) is adopted from the previous work (Sec. 2.3) and the oxygen diffusivity ($D_O$) (Sec. 2.2) is obtained from AIMD simulations. Figure 3 shows the calculated oxygen conductivities of BCFZY materials compared with known experimental values. It was observed that Y doping reduces the oxygen conductivity, which is consistent with experimental results. The calculated activation energies, obtained from Arrhenius fits to $\sigma T$, are 0.25 ± 0.001 eV for BCFZ, 0.27 ± 0.001 eV for BCFZY0.1 and 0.36 ± 0.001 eV for BCFY. While both modelling and experiments show the expected decrease in the conductivity with increasing Y content, they have quite different slopes and magnitudes of oxygen conductivity at the same temperatures. Further, the experiments show some significant deviations from Arrhenius behavior. This combination of differences leads to deviations of model-predicted and experimentally measured conductivities of over an order of magnitude by about 830K, the lowest temperatures measured. Possibly, this discrepancy is mainly caused by the role of grain boundaries, a hypothesis that is discussed in greater detail in Sec. 3.2.

### 3.2. Discussions of model discrepancies with experiment

#### 3.2.1. Effect of grain boundaries vs. Alternative Hypotheses for Discrepancies

The origin of the discrepancies in linearity, slopes, and values observed in Figure 3 are not rigorously established, but we speculate that they can be attributed at least in part to the effect of grain boundaries, which often slow down oxygen conductivity in polycrystalline samples.[52–54] Several possible alternative explanations were examined and excluded, lending at least some support to the grain boundary hypothesis. First, potential errors in the calculation of oxygen diffusion were considered, particularly the possible influence of our approximate treatment of the magnetism (see SI Sec. 2). A comparison of AIMD simulations with different magnetic configurations, where the initial magnetic moments were set on individual atoms and the total magnetic moment of the system was fixed throughout, showed that the resulting fluctuations in the magnetic moments of transition metal atoms in BCFZ had a negligible effect on the calculated oxygen diffusivity. This indicates that the magnetic treatment is not a significant source of error. Additionally, we evaluated whether surface exchange might be influencing the experimental BCFZY conductivity measurements, based on the analysis reported by Duffy et al [25] (see SI Sec. 3). Their application of the Wagner equation with and without correction for surface exchange indicated that surface kinetics could cause up to a 40% deviation in BCFZ and less than 5% in BCFY. However, this level of deviation is still too small to account for the much larger discrepancy, which often exceeds an order of magnitude, between experimental and modeled conductivities. Together, these investigations indicate that neither magnetic treatment nor surface exchange is the primary source of deviation, and thus, by way of elimination of at least some alternatives, supports the hypothesis that grain boundary diffusion may be the dominant factor influencing the experimental results.

#### 3.2.2. Modeling effects of grain boundaries on oxygen conductivity

We hypothesize the dominant reason behind the different behavior in oxygen conductivity seen in the polycrystalline experiments compared to the single crystal model predictions is grain boundary effects. Grain boundaries in BCFZ are expected to reduce oxygen conductivity. It is reported that activation energies for oxygen permeation through 1 mm thick $BaCo_{0.7}Fe_{0.24}Zr_{0.06}O_{3-\delta}$ membranes, sintered at 1110 °C, decrease as dwell time (and thus grain size) increases. In the 750 to 800 °C range, the activation energies drop from 0.590 eV (5 h dwelling) to 0.505 eV (50 h dwelling) and 0.436 eV (100 h dwelling); in the 800 to 925 °C range, it falls from 0.352 eV (5 h dwelling) to 0.324 eV (50 h dwelling) and 0.301 eV (100 h dwelling). The monotonic decrease in the activation energies with increasing grain size indicates that the grain boundaries in $BaCo_{0.7}Fe_{0.24}Zr_{0.06}O_{3-\delta}$ impede oxygen transport, so larger grains promote oxygen permeation. [55] Numerous examples in the literature also show that oxygen conductors often exhibit much lower grain boundary conductivity than bulk conductivity. For example, $K_{0.25}(Ca_{0.5}Nb_{0.5})O_{2.75}$ (BKCN25), a perovskite material intended for use as an electrolyte in solid oxide fuel cells (SOFCs), exhibits a significant discrepancy in oxygen conductivity between the bulk and grain boundaries.[56] The activation energy for oxygen conductivity increases from 0.26 eV in the bulk to 1.35 eV at the grain boundaries. Similarly, for $La_{0.5}Sr_{0.5}FeO_{3-\delta}$ (LSF), it has been reported that the activation energy for oxygen conductivity decreases as the grain size increases, corresponding to 2.52 eV, 2.29 eV and 1.95 eV for grain sizes of 0.20 $\mu m$, 0.63 $\mu m$ and 1.43 $\mu m$, respectively. [57] For $La_{0.84}Sr_{0.16}CoO_{3-\delta}$ (LSC16), the activation energy for oxygen diffusivity in a single-crystal sample (1.31 eV) is lower than that in a polycrystalline sample (1.75 eV), implying the blocking effect of grain boundaries [58] Likewise, studies on for $Y_2O_3$-doped $CeO_2$, which is one of the most important oxygen ion-conducting solid electrolytes, show that the activation energy for oxygen conductivity increases from 0.70 eV in the bulk to 1.36 eV at the grain boundaries.[59] These results suggest that 0.5-1 eV increases in activation energies for oxygen conduction from bulk to grain boundary are quite reasonable.

In BCFZ, the experimental activation energy for oxygen conductivity (about 0.675 eV) is about 0.5 eV larger than that predicted from our bulk materials modelling (0.155 eV). If we assume that our modeling is correctly predicting the single crystal bulk conductivity, and that the transport in the BCFZ experiments is dominated by grain boundaries, then these values suggest the grain boundaries have about 0.5 eV higher barriers than the bulk, a reasonable value consistent with many previous studies.

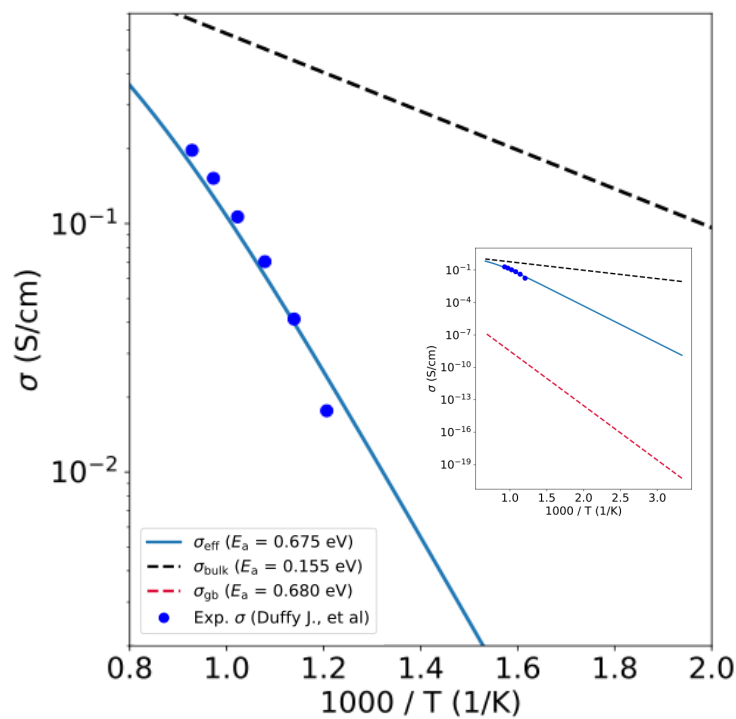

Figure 4. Grain boundary oxygen conductivity ($\sigma_{gb}$) calculated using the series diffusion model, with bulk oxygen conductivity ($\sigma_{bulk}$) taken from our model and effective oxygen conductivity ($\sigma_{eff}$) taken from experimental data.

To make this reasoning more quantitative, we used a simple series model which describes the effective ionic conductivity ($\sigma_{eff}$) in terms of bulk ($\sigma_b$) and grain boundary ($\sigma_{gb}$) conductivities by considering them to occur in series, yielding the equation:

$$\sigma_{eff} = \frac{(L+d)^2}{\frac{L^2}{\sigma_b}+\frac{d^2}{\sigma_{gb}}} \quad (Equation\ 2)$$

Here, L is grain size and d is grain boundary thickness. We assumed $L = 1000d$. In our analysis, we treated the oxygen conductivity predicted by our model as the bulk value, and the experimental reference as the effective conductivity. For BCFZ, our model yields a bulk oxygen conductivity ($\sigma_{bulk}$) with an activation energy of 0.155 eV and a pre-exponential factor of 3.50 S/cm. The experimental oxygen conductivity ($\sigma_{eff}$) shows a much higher activation energy of 0.675 eV and a pre-exponential factor of $3.04 \times 10^2$ S/cm. Using the series diffusion model and fitting to the experimental conductivities, we estimate the activation energy for oxygen transport at the grain boundary to be 0.675 eV, indicating that grain boundary conductivity dominates the effective oxygen transport, as shown in Figure 4. The corresponding pre-exponential factor is calculated to be $3.49 \times 10^{-4}$ S/cm, which is reasonable and consistent with literature reports that prefactors for grain boundary oxygen conductivities are typically two to four orders of magnitude lower than bulk values. [60,61] Therefore, our model is likely reasonably accurate for the single crystal, and experiments to date are strongly influenced by a much slower grain boundary conductivity. These findings indicate that materials processing such as sintering protocols to increase the grain size and alter the grain boundary structure could yield significantly improved oxygen conduction.

## 4. Conclusion

In this study, we investigated the effect of Y doping on the oxygen conductivity of Ba (Fe, Co, Zr, Y) O₃ (BCFZY) materials, which are promising candidates for cathodes in protonic ceramic fuel cells (PCFCs). Using AIMD simulations, we found that Y doping leads to a slight decrease in oxygen tracer diffusivity. The migration energy for oxygen diffusion shows no statistically significant change between BCFZ (0.25 eV) and BCFZY0.1 (0.27 eV), but increased significantly in BCFY (0.36 eV). These barrier effects will tend to slow oxygen diffusion with increasing Y content. However, reference data of oxygen nonstoichiometry revealed that Y doping increases the oxygen vacancy concentration due to its lower valence compared to Zr, which will tend to increase oxygen diffusion with Y content. Calculations using the Nernst-Einstein relation showed that Y doping overall reduces the oxygen conductivity, which aligns with experimental data. This reduction arises primarily from a decrease in oxygen vacancy diffusivity ($D_{vac}$), which dominate over the increases in $c_{vac}$. A significant discrepancy was observed between the calculated and experimental oxygen conductivities, with the experimental values being an order of magnitude lower and much less Arrhenius-like in their temperature dependence. We hypothesized that this discrepancy may arise from slower oxygen diffusion at grain boundaries. To explore this possibility, we performed a model fit assuming the presence of a slower grain boundary pathway. The fit suggested that a grain boundary with a pre-exponential factor about four orders of magnitude smaller than our calculated bulk value, and an activation energy about 0.5 eV higher than our calculated bulk value, could produce the effective conductivity that matches the experimentally observed oxygen conductivity. This result is consistent with pre-factor and barrier differences of other perovskites reported in the literature, lending some support to our hypothesis. Our results highlight the potentially very significant role of grain boundaries in limiting oxygen transport in these materials. The results suggest that tailoring the microstructure through methods such as promoting grain growth, adjusting doping concentrations and grain boundary segregation, and generally engineering grain boundary structure and chemistry could be important for improving cathode performance. The insights gained from this study not only clarify the mechanism of Y doping effects on bulk transport but also provide guidance on microstructural issues critical for designing next-generation PCFC cathodes with improved efficiency and stability.

## Acknowledgements

This work was supported by the U.S. Department of Energy's National Energy Technology Laboratory (NETL), in part through a site support contract. Neither the United States Government nor any agency thereof, nor any of their employees, nor the support contractor or its employees, makes any warranty, express or implied, or assumes any legal liability or responsibility for the accuracy, completeness, or usefulness of any information, apparatus, product, or process disclosed, or represents that its use would not infringe privately owned rights. Reference to any specific commercial product, process, or service by trade name, trademark, manufacturer, or otherwise does not constitute or imply its endorsement, recommendation, or favoring by the United States Government or any agency thereof. The views and opinions expressed herein are those of the authors and do not necessarily reflect those of the United States Government or any agency thereof. Computations were performed on the Stampede3 system at the Texas Advanced Computing Center (TACC), The University of Texas at Austin, under allocation TG-MAT240071 from the Advanced Cyberinfrastructure Coordination Ecosystem: Services & Support (ACCESS) program, supported by National Science Foundation grants 2138259, 2138286, 2138307, 2137603, and 2138296. [62] Additional computational resources were provided by the Center for High Throughput Computing at the University of Wisconsin–Madison. [63] C.K., R.J., and D.M. thank Dr. Yueh-Lin Lee and Dr. Jun Meng for their advice on DFT calculations and for valuable discussions throughout this work.

## Conflict of Interest

The authors declare no conflicts or interests.

## Data Availability Statement

The data that support the findings of this study are openly available in Figshare at https://doi.org/10.6084/m9.figshare.30276166.v1, reference number [64].

**Supplementary information for**

Effects of Yttrium doping on Oxygen Conductivity in Ba (Fe, Co, Zr, Y) $O_{3-\delta}$ Cathode Materials for Proton Ceramic Fuel Cells

Chiyoung Kim[a], Ryan Jacobs[a], Jack H. Duffy[b], Kyle S. Brinkman[b], Harry W. Abernathy[c], Dane Morgan[a]

a. Department of Materials Science and Engineering, University of Wisconsin – Madison, Madison, Wisconsin 54706, United States
b. Department of Materials Science and Engineering, Clemson University, Clemson, South Carolina 29634, United States
c. National Energy Technology Laboratory, United States Department of Energy, Morgantown, West Virginia 26507, United States

# 1. Oxygen tracer diffusivity and migration energy results from MD calculations for each configuration and cell size.

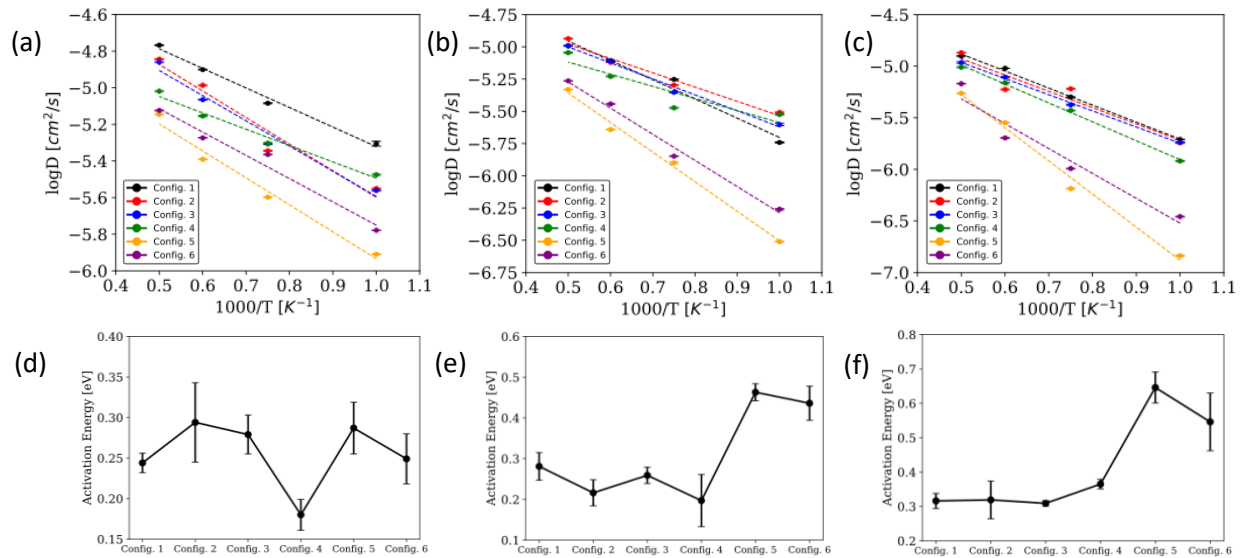

Figure S1. Tracer diffusivity of oxygen atom (a) and migration energy (d) for oxygen diffusion in BCFZ; tracer diffusivity (b) and migration energy (e) in BCFZY0.1; and tracer diffusivity (c) and migration energy (f) in BCFY – all with six different B-site configurations

Error bars for the oxygen diffusion coefficients were obtained from the standard error of the slope of a linear fit to the time-averaged mean square displacement (TAMSD) as a function of time, using non-linear least-squares regression (curve_fit in SciPy). The diffusion coefficient was calculated from the fitted slope based on the Einstein relation, $D = \frac{slope}{6}$. The uncertainty in the oxygen migration energy ($E_{mig}$) was then determined by propagating these slope-derived uncertainties through a weighted Arrhenius fit of lnD versus 1/T, where the diffusion coefficient uncertainties were used as weights to appropriately account for confidence in each data point.

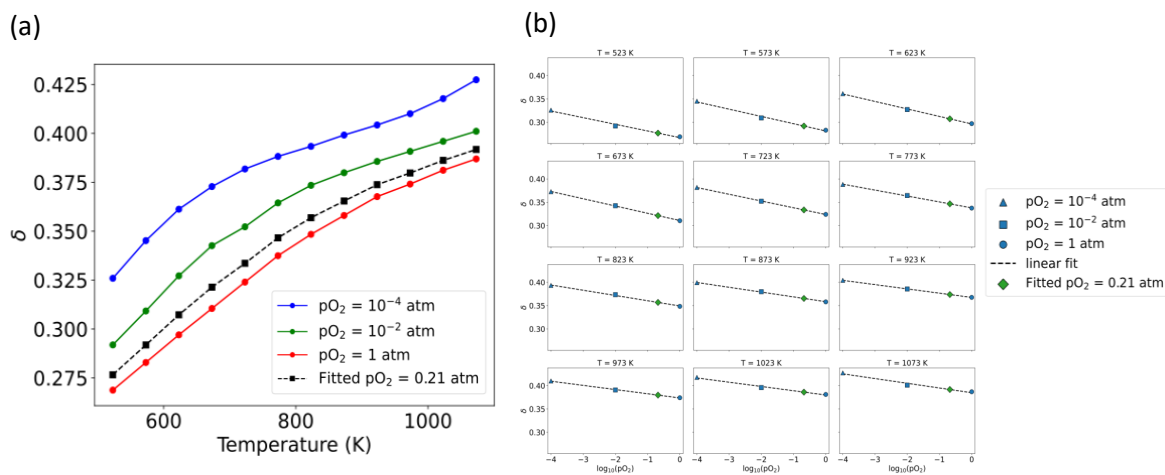

Figure S2. (a) Oxygen non-stoichiometry ($\delta$), versus temperature T(K) at pO$_2$ = 10$^{-4}$, 10$^{-2}$, and 1atm. The dashed black curve shows $\delta$ at air, pO$_2$ = 0.21 atm, obtained at each T by linear regression of $\delta$ vs. ln(pO$_2$) using the three measured pO$_2$ values and then evaluating at 0.21 atm. (b) For each T, $\delta$ as a function of log(pO$_2$) with the corresponding linear fit (dashed line). The fitted point at 0.21 atm is highlighted (green diamond). The fitting form is $\delta(T, pO_2) = \alpha(T) + \beta(T)\ln(pO_2)$

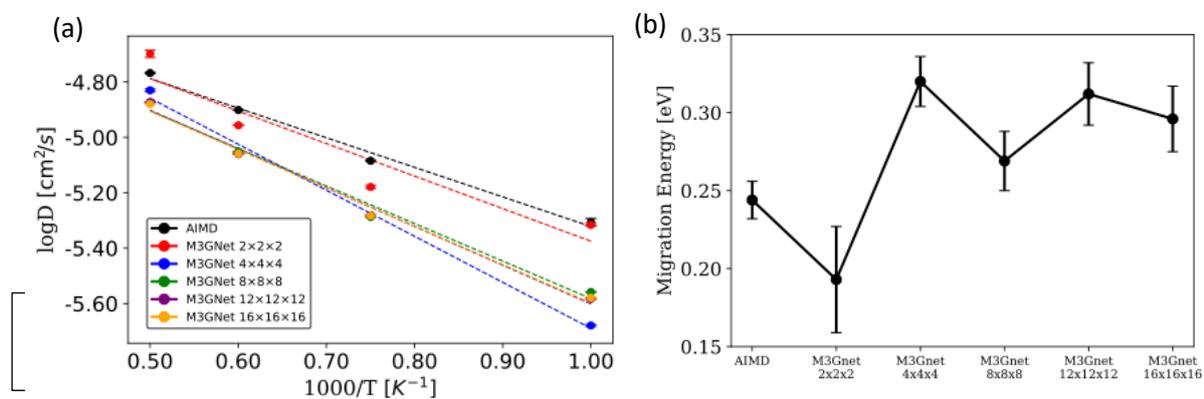

Figure S3. Tracer diffusivity (a) and $E_{mig}$ (b) of oxygen atom in BCFZ with different cell size

Oxygen diffusion in BCFZY is effectively size-converged, with minimal impact of supercell size on diffusivity or migration energy.

## 2. Possible error in the calculation of oxygen diffusion due to magnetic approach

It is expected that the BCFZY system is in a paramagnetic state at the temperatures of the experiments (Duffy, J. et al[40]). Due to the limitations of the ab initio methods, it is not practical to simulate this magnetic state directly. In the results shown in Figure  (Sec. 2.), the system was initialized in a ferromagnetic configuration (see the "Methods" section), where all $Fe^{3+}$ and $Co^{3+}$ ions were set in high-spin states with parallel alignment. The initial magnetic moments were assigned as 5 $\mu_B$ for $Fe^{3+}$ and 4 $\mu_B$ for $Co^{3+}$, corresponding to $d^5$ and $d^6$ electron configurations, respectively. The total magnetic moment of the supercell was fixed throughout the AIMD simulations, and the magnetic state was allowed to evolve as determined by the simulation. This setup allows for magnetic fluctuations that might occur due to vibrations but also may create unphysical magnetic fluctuations. Therefore, here we investigated whether the fluctuation in the magnetic moment of each transition metal atom in BCFZ unit cell during AIMD simulation might impact the diffusion and potentially cause the error in the diffusivity calculation. Figure S4 shows the magnetic moment ($\mu_B$) of transition metal atoms for 10ps of AIMD simulation at 1000 K, 1333 K, 1666 K and 2000 K with the same setup used for Figure . In this case, fluctuation of the magnetic moment becomes significant at the higher temperature. Also, the fluctuation is more severe for the cobalt atoms than iron atoms. The varying moments of $Co^{3+}$ suggested checking the intermediate spin arrangement for $Co^{3+}$ (3 $\mu_B$ / Co atom) was also important. This spin could be stabilized by assigning an initial magnetic moment of 3 $\mu_B$ for each $Co^{3+}$ ion (Figure S5). We therefore ran the same AIMD as before except assigning the initial magnetic moment of 3 $\mu_B$ for each $Co^{3+}$ ion and constraining the total moment appropriately. In this case, more significant fluctuation of magnetic moment than in Figure S4 was observed in all temperature ranges. The oxygen tracer diffusivities were calculated from the AIMD calculation of the two assigned initial magnetic moment cases (with 3 $\mu_B$ / Co atom and 4 $\mu_B$ / Co atom). (Figure S6) It was observed that the difference of oxygen diffusivities between each case is within the overall uncertainties of the diffusion coefficients. From comparing oxygen diffusivities of the two cases, it was observed that the discrepancy caused by the fluctuation of magnetic moment is negligible on the scale of uncertainties in this work and not the cause of the discrepancies between predicted and measured oxygen conductivity (Figure 3).

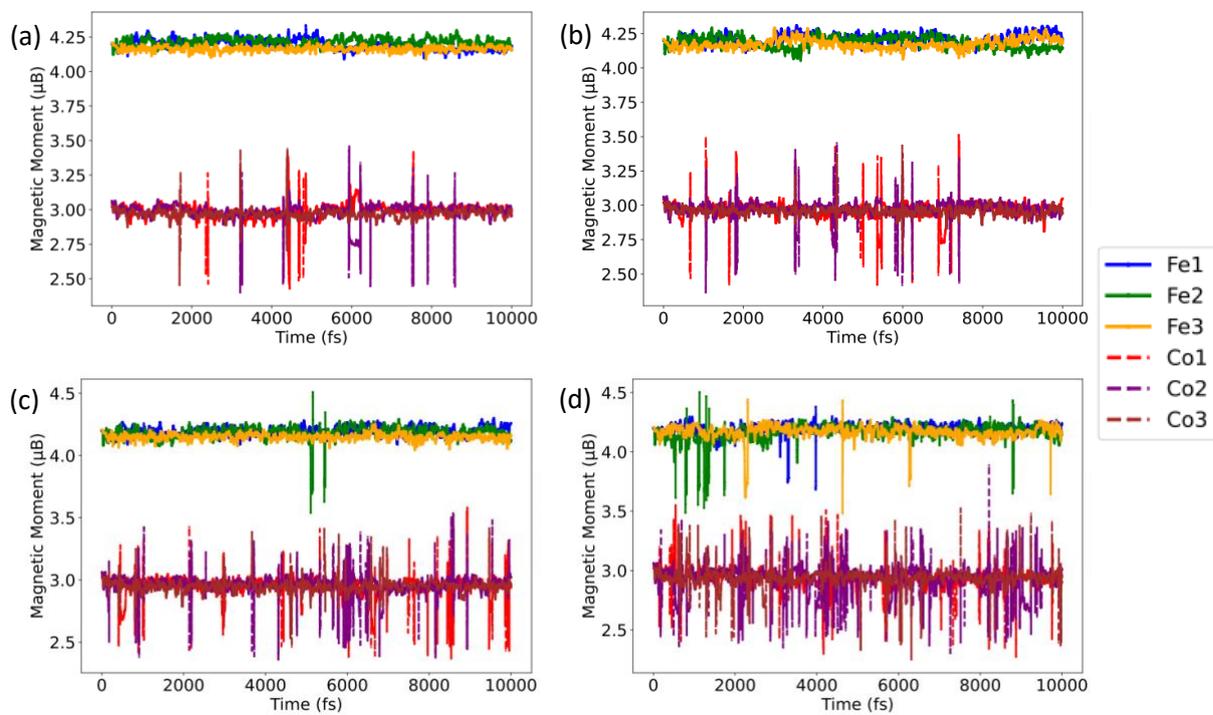

Figure S4. Magnetic moment fluctuation in BCFZ during AIMD simulation with a assigned initial magnetic moment of $4\,\mu_B$ per each $Co^{3+}$ ion. (a) 1000K, (b) 1333K, (c) 1666K and (d) 2000K.

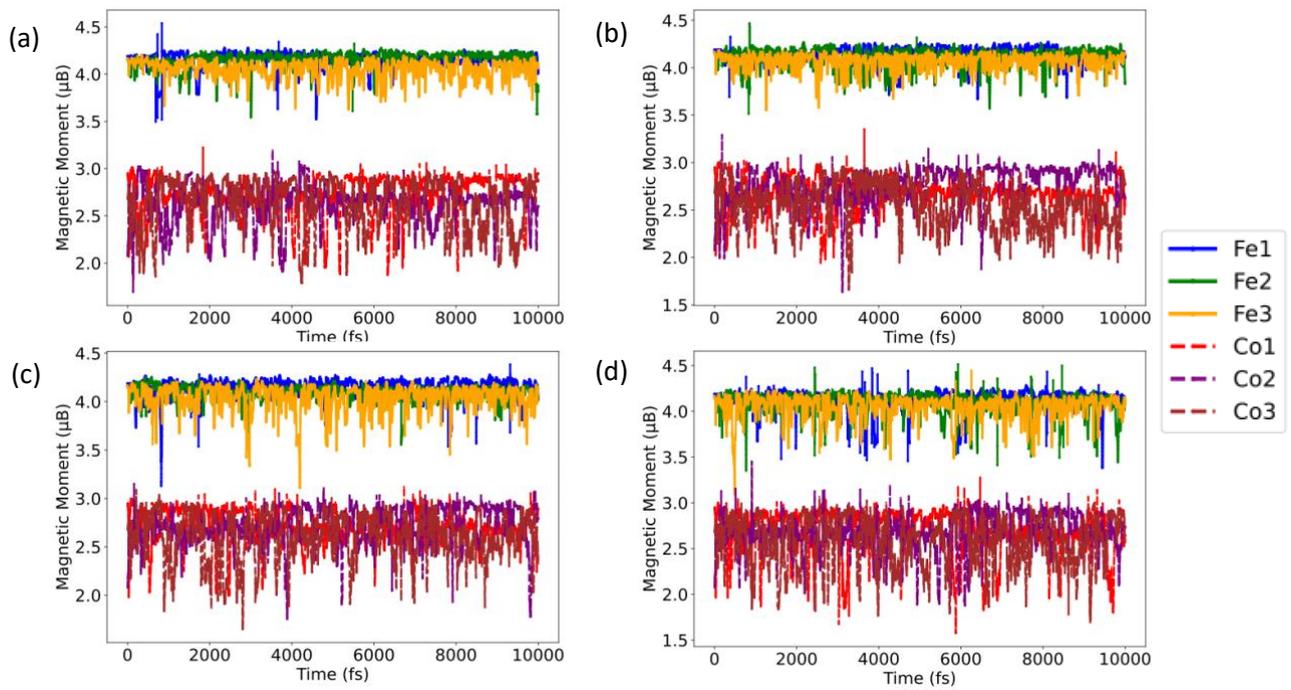

Figure S5. Magnetic moment fluctuation in BCFZ during AIMD simulation with a assigned initial magnetic moment of $3\ \mu_B$ per each $Co^{3+}$ ion. (a) 1000K, (b) 1333K, (c) 1666K and (d) 2000K.

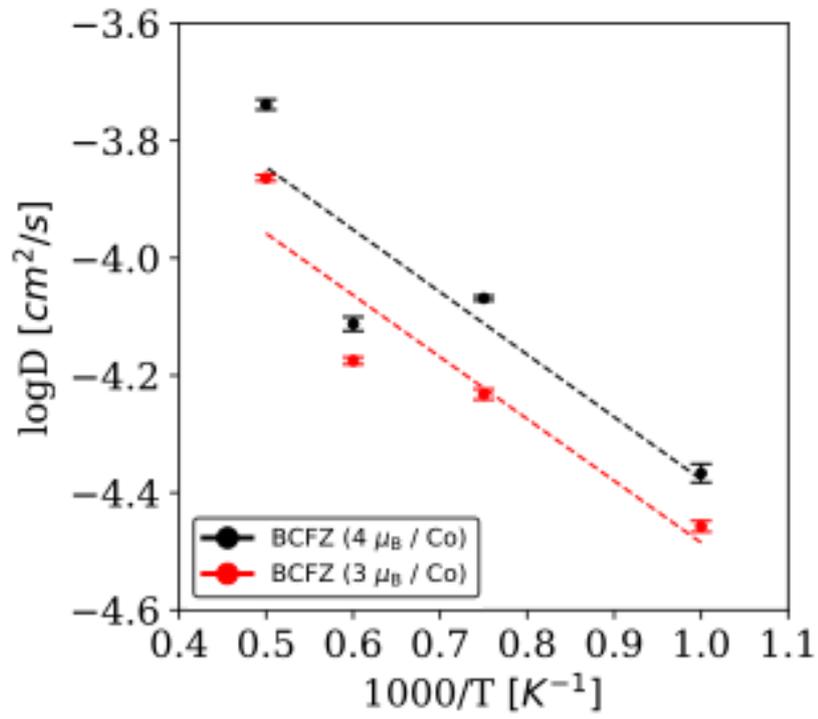

Figure S6. Tracer diffusivity of oxygen atom in BCFZ with fixed magnetic moment as $4\ \mu_B$ and $3\ \mu_B$ per each $Co^{3+}$ ion.

## 3. Influence of surface exchange rate in experimental BCFZY conductivity measurements

To ensure that the comparison of oxygen conductivities between experiment and simulation results is on the same basis, it is essential to assess whether the experimental measurements are governed primarily by bulk diffusion or are significantly influenced by surface exchange kinetics. In gas permeation experiments, the oxygen flux across dense membranes can be limited by both bulk ionic transport and surface exchange reactions. The relative contributions of these processes depend on the characteristic thickness ($L_C$) of the material, which defines the length scale below which surface kinetics begin to significantly influence the overall transport behavior. [65]

The oxygen permeation flux $J_{O_2}$ in such membranes is often estimated using the Wagner equation, which, under the assumption of purely bulk-limited transport, is given as:

$$J_{O_2} = -\frac{RT}{16F^2L}\sigma_{ion}\ln\left(\frac{P_{O_2}''}{P_{O_2}'}\right) \quad \text{(Equation S1)}$$

Here, R is the universal gas constant, T is the temperature, F is the Faraday constant, L is the membrane thickness, $\sigma_{ion}$ is the ionic conductivity, and $P_{O_2}'$, $P_{O_2}''$ are the oxygen partial pressures on each side of the membrane. However, when surface exchange resistance is non-negligible, a correction factor is introduced to account for its contribution, yielding a modified form of the Wagner equation:

$$J_{O_2} = -\frac{1}{1+2\left(\frac{L_C}{L}\right)}\frac{RT}{16F^2L}\sigma_{ion}\ln\left(\frac{P_{O_2}''}{P_{O_2}'}\right) \quad \text{(Equation S2)}$$

In this form, $L_C$ is the material's characteristic thickness, and the prefactor accounts for the additional resistance introduced by surface exchange limitations. J. Duffy et al. [25] applied this correction to interpret their gas permeation measurements on BCFZYx samples with a membrane thickness (L) of 1 mm. They reported that the characteristic thickness ($L_C$) decreased significantly with increasing Y content: 209 μm for BCFZ, 84 μm for BCFZY0.1, and 21 μm for BCFY, all evaluated at 600 °C. Due to this variation, the influence of surface exchange on the extracted oxygen conductivity also varied. According to their analysis, correcting for surface exchange led to a deviation of approximately 40 % in the calculated conductivity for BCFZ and only about 4 % for BCFY. These results suggest that oxygen transport in BCFY is predominantly governed by bulk diffusion, whereas surface exchange kinetics play a more significant role in BCFZ. While surface exchange resistance can introduce meaningful deviations in the extracted conductivity values, even in the case of BCFZ, which is the most affected composition, the deviation reported by J. Duffy et al. [25] remains under 50 %. This is insufficient to account for a significant fraction of the discrepancy observed between our simulated bulk oxygen conductivity and the experimental data, which, in some cases, differs by more than an order of magnitude.

BCFZ

| T(K) | c$_{vac}$_BCFZ [mol/f.u.] | λ_BCFZ | D$_{vac}$_BCFZ [cm$^2$/s] | D$_O$_BCFZ [cm$^2$/s] | σ_BCFZ [S/cm] | Tσ_BCFZ [S·K/cm] |
|---|---|---|---|---|---|---|
| 523 | 0.2766 | 0.1016 | 1.62E-06 | 1.64E-07 | 0.07939 | 41.52 |
| 573 | 0.2918 | 0.1078 | 2.50E-06 | 2.69E-07 | 0.118 | 67.63 |
| 623 | 0.3072 | 0.1141 | 3.60E-06 | 4.10E-07 | 0.1646 | 102.5 |
| 673 | 0.3214 | 0.12 | 4.90E-06 | 5.88E-07 | 0.2173 | 146.3 |
| 723 | 0.3336 | 0.1251 | 6.41E-06 | 8.02E-07 | 0.2744 | 198.4 |
| 773 | 0.3466 | 0.1306 | 8.09E-06 | 1.06E-06 | 0.3366 | 260.2 |
| 823 | 0.3569 | 0.135 | 9.92E-06 | 1.34E-06 | 0.3994 | 328.7 |
| 873 | 0.3654 | 0.1387 | 1.19E-05 | 1.65E-06 | 0.4621 | 403.4 |
| 923 | 0.3738 | 0.1423 | 1.40E-05 | 1.99E-06 | 0.5254 | 484.9 |
| 973 | 0.3798 | 0.1449 | 1.62E-05 | 2.34E-06 | 0.5853 | 569.5 |
| 1000 | 0.3837 | 0.1467 | 1.74E-05 | 2.55E-06 | 0.6185 | 618.5 |
| 1333 | 0.4142 | 0.1602 | 3.33E-05 | 5.33E-06 | 0.9598 | 1279 |
| 1666 | 0.429 | 0.1669 | 4.92E-05 | 8.20E-06 | 1.175 | 1958 |
| 2000 | 0.4361 | 0.1701 | 6.38E-05 | 1.09E-05 | 1.292 | 2583 |

BCFZY

| T(K) | c$_{vac}$_BCFZ [mol/f.u.] | λ_BCFZ | D$_{vac}$_BCFZ [cm$^2$/s] | D$_O$_BCFZ [cm$^2$/s] | σ_BCFZ [S/cm] | Tσ_BCFZ [S·K/cm] |
|---|---|---|---|---|---|---|
| 523 | 0.3266 | 0.1222 | 7.43E-07 | 9.07E-08 | 0.04198 | 21.96 |
| 573 | 0.3418 | 0.1286 | 1.20E-06 | 1.54E-07 | 0.06466 | 37.05 |
| 623 | 0.3572 | 0.1352 | 1.79E-06 | 2.42E-07 | 0.09282 | 57.83 |
| 673 | 0.3714 | 0.1413 | 2.52E-06 | 3.56E-07 | 0.1257 | 84.58 |
| 723 | 0.3836 | 0.1466 | 3.38E-06 | 4.95E-07 | 0.1622 | 117.2 |
| 773 | 0.3966 | 0.1523 | 4.36E-06 | 6.65E-07 | 0.2026 | 156.6 |
| 823 | 0.4069 | 0.1569 | 5.46E-06 | 8.57E-07 | 0.2445 | 201.2 |
| 873 | 0.4154 | 0.1607 | 6.67E-06 | 1.07E-06 | 0.2872 | 250.7 |
| 923 | 0.4238 | 0.1645 | 7.97E-06 | 1.31E-06 | 0.331 | 305.5 |
| 973 | 0.4298 | 0.1672 | 9.34E-06 | 1.56E-06 | 0.3735 | 363.4 |
| 1000 | 0.4337 | 0.169 | 1.01E-05 | 1.71E-06 | 0.397 | 397 |
| 1333 | 0.4642 | 0.1831 | 2.07E-05 | 3.79E-06 | 0.6519 | 869 |
| 1666 | 0.479 | 0.19 | 3.18E-05 | 6.04E-06 | 0.8268 | 1377 |
| 2000 | 0.4861 | 0.1934 | 4.23E-05 | 8.19E-06 | 0.9314 | 1863 |

BCFY

| T(K) | $c_{vac}$_BCFZ [mol/f.u.] | λ_BCFZ | $D_{vac}$_BCFZ [cm$^2$/s] | $D_O$_BCFZ [cm$^2$/s] | σ_BCFZ [S/cm] | Tσ_BCFZ [S·K/cm] |
|---|---|---|---|---|---|---|
| 523 | 0.3766 | 0.1436 | 1.72E-07 | 2.47E-08 | 0.01091 | 5.708 |
| 573 | 0.3918 | 0.1502 | 3.31E-07 | 4.97E-08 | 0.01994 | 11.43 |
| 623 | 0.4072 | 0.1571 | 5.73E-07 | 9.00E-08 | 0.03303 | 20.58 |
| 673 | 0.4214 | 0.1634 | 9.15E-07 | 1.49E-07 | 0.05052 | 34 |
| 723 | 0.4336 | 0.1689 | 1.37E-06 | 2.31E-07 | 0.07241 | 52.35 |
| 773 | 0.4466 | 0.1749 | 1.94E-06 | 3.40E-07 | 0.0991 | 76.61 |
| 823 | 0.4569 | 0.1796 | 2.65E-06 | 4.76E-07 | 0.1296 | 106.7 |
| 873 | 0.4654 | 0.1836 | 3.48E-06 | 6.39E-07 | 0.1636 | 142.8 |
| 923 | 0.4738 | 0.1875 | 4.44E-06 | 8.32E-07 | 0.2009 | 185.4 |
| 973 | 0.4798 | 0.1904 | 5.52E-06 | 1.05E-06 | 0.2401 | 233.6 |
| 1000 | 0.4837 | 0.1922 | 6.16E-06 | 1.18E-06 | 0.2626 | 262.6 |
| 1333 | 0.5142 | 0.2069 | 1.64E-05 | 3.39E-06 | 0.5582 | 744.1 |
| 1666 | 0.529 | 0.2141 | 2.95E-05 | 6.33E-06 | 0.8275 | 1379 |
| 2000 | 0.5361 | 0.2176 | 4.38E-05 | 9.53E-06 | 1.035 | 2071 |

Table S1 Summary of the parameters used: oxygen vacancy concentration ($c_{vac}$), oxygen tracer correlation factor (Λ), oxygen vacancy diffusivity ($D_{vac}$), oxygen ion diffusivity ($D_O$), ionic conductivity (σ) and temperature-normalized conductivity (σT) for BCFZ, BCFZY and BCFY